\documentclass[aps,superscriptaddress,prd,showpacs,12pt]{revtex4}
\usepackage{graphicx}
\usepackage{amsmath}
\usepackage{epsfig}
\usepackage{amssymb}

\begin{document}

\title{\bf\boldmath Search for the $\eta^{\prime}\to e^+e^-$ decay with
the SND detector}
\author{M.N.~Achasov}
\affiliation{Budker Institute of Nuclear Physics, SB RAS, Novosibirsk, 630090, Russia}
\affiliation{Novosibirsk State University, Novosibirsk 630090, Russia}
\author{V.M.~Aulchenko}
\affiliation{Budker Institute of Nuclear Physics, SB RAS, Novosibirsk, 630090, Russia}
\affiliation{Novosibirsk State University, Novosibirsk 630090, Russia}
\author{A.Yu.~Barnyakov}
\affiliation{Budker Institute of Nuclear Physics, SB RAS, Novosibirsk, 630090, Russia}
\affiliation{Novosibirsk State University, Novosibirsk 630090, Russia}
\author{K.I.~Beloborodov}
\affiliation{Budker Institute of Nuclear Physics, SB RAS, Novosibirsk, 630090, Russia}
\affiliation{Novosibirsk State University, Novosibirsk 630090, Russia}
\author{A.V.~Berdyugin}
\affiliation{Budker Institute of Nuclear Physics, SB RAS, Novosibirsk, 630090, Russia}
\affiliation{Novosibirsk State University, Novosibirsk 630090, Russia}
\author{D.E.~Berkaev}
\affiliation{Budker Institute of Nuclear Physics, SB RAS, Novosibirsk, 630090, Russia}
\affiliation{Novosibirsk State University, Novosibirsk 630090, Russia}
\author{A.G.~Bogdanchikov}
\affiliation{Budker Institute of Nuclear Physics, SB RAS, Novosibirsk, 630090, Russia}
\author{A.A.~Botov}
\affiliation{Budker Institute of Nuclear Physics, SB RAS, Novosibirsk, 630090, Russia}
\author{T.V.~Dimova}
\email[]{baiert@inp.nsk.su}
\affiliation{Budker Institute of Nuclear Physics, SB RAS, Novosibirsk, 630090, Russia}
\affiliation{Novosibirsk State University, Novosibirsk 630090, Russia}
\author{V.P.~Druzhinin}
\affiliation{Budker Institute of Nuclear Physics, SB RAS, Novosibirsk, 630090, Russia}
\affiliation{Novosibirsk State University, Novosibirsk 630090, Russia}
\author{V.B.~Golubev}
\affiliation{Budker Institute of Nuclear Physics, SB RAS, Novosibirsk, 630090, Russia}
\affiliation{Novosibirsk State University, Novosibirsk 630090, Russia}
\author{ L.~V.~Kardapoltsev}
\affiliation{Budker Institute of Nuclear Physics, SB RAS, Novosibirsk, 630090, Russia}
\affiliation{Novosibirsk State University, Novosibirsk 630090, Russia}
\author{A.S.~Kasaev}
\affiliation{Budker Institute of Nuclear Physics, SB RAS, Novosibirsk, 630090, Russia}
\author{A.G.~Kharlamov}
\affiliation{Budker Institute of Nuclear Physics, SB RAS, Novosibirsk, 630090, Russia}
\affiliation{Novosibirsk State University, Novosibirsk 630090, Russia}
\author{A.N.~Kirpotin}
\affiliation{Budker Institute of Nuclear Physics, SB RAS, Novosibirsk, 630090, Russia}
\author{D.P.~Kovrizhin}
\affiliation{Budker Institute of Nuclear Physics, SB RAS, Novosibirsk, 630090, Russia}
\affiliation{Novosibirsk State University, Novosibirsk 630090, Russia}
\author{I.A.~Koop}
\affiliation{Budker Institute of Nuclear Physics, SB RAS, Novosibirsk, 630090, Russia}
\affiliation{Novosibirsk State University, Novosibirsk 630090, Russia}
\affiliation{Novosibirsk State Technical University, Novosibirsk, 630092, Russia}
\author{A.A.~Korol}
\affiliation{Budker Institute of Nuclear Physics, SB RAS, Novosibirsk, 630090, Russia}
\affiliation{Novosibirsk State University, Novosibirsk 630090, Russia}
\author{S.V.~Koshuba}
\affiliation{Budker Institute of Nuclear Physics, SB RAS, Novosibirsk, 630090, Russia}
\affiliation{Novosibirsk State University, Novosibirsk 630090, Russia}
\author{A.S.~Kupich}
\affiliation{Budker Institute of Nuclear Physics, SB RAS, Novosibirsk, 630090, Russia}
\affiliation{Novosibirsk State University, Novosibirsk 630090, Russia}
\author{K.A.~Martin}
\affiliation{Budker Institute of Nuclear Physics, SB RAS, Novosibirsk, 630090, Russia}
\affiliation{Novosibirsk State University, Novosibirsk 630090, Russia}
\author{N.Yu.~Muchnoi}
\affiliation{Budker Institute of Nuclear Physics, SB RAS, Novosibirsk, 630090, Russia}
\affiliation{Novosibirsk State University, Novosibirsk 630090, Russia}
\author{A.E.~Obrazovsky}
\affiliation{Budker Institute of Nuclear Physics, SB RAS, Novosibirsk, 630090, Russia}
\author{A.V.~Otboev}
\affiliation{Budker Institute of Nuclear Physics, SB RAS, Novosibirsk, 630090, Russia}
\affiliation{Novosibirsk State University, Novosibirsk 630090, Russia}
\author{E.V.~Pakhtusova}
\affiliation{Budker Institute of Nuclear Physics, SB RAS, Novosibirsk, 630090, Russia}
\author{A.I.~Senchenko}
\affiliation{Budker Institute of Nuclear Physics, SB RAS, Novosibirsk, 630090, Russia}
\affiliation{Novosibirsk State University, Novosibirsk 630090, Russia}
\author{S.I.~Serednyakov}
\affiliation{Budker Institute of Nuclear Physics, SB RAS, Novosibirsk, 630090, Russia}
\affiliation{Novosibirsk State University, Novosibirsk 630090, Russia}
\author{P.Yu.~Shatunov}
\affiliation{Budker Institute of Nuclear Physics, SB RAS, Novosibirsk, 630090, Russia}
\affiliation{Novosibirsk State University, Novosibirsk 630090, Russia}
\author{Yu.M.~Shatunov}
\affiliation{Budker Institute of Nuclear Physics, SB RAS, Novosibirsk, 630090, Russia}
\affiliation{Novosibirsk State University, Novosibirsk 630090, Russia}
\author{D.A.~Shtol}
\affiliation{Budker Institute of Nuclear Physics, SB RAS, Novosibirsk, 630090, Russia}
\author{D.B.~Shwartz}
\affiliation{Budker Institute of Nuclear Physics, SB RAS, Novosibirsk, 630090, Russia}
\affiliation{Novosibirsk State University, Novosibirsk 630090, Russia}
\author{Z.K.~Silagadze}
\affiliation{Budker Institute of Nuclear Physics, SB RAS, Novosibirsk, 630090, Russia}
\affiliation{Novosibirsk State University, Novosibirsk 630090, Russia}
\author{I.K.~Surin}
\affiliation{Budker Institute of Nuclear Physics, SB RAS, Novosibirsk, 630090, Russia}
\author{Yu.A.~Tikhonov}
\affiliation{Budker Institute of Nuclear Physics, SB RAS, Novosibirsk, 630090, Russia}
\affiliation{Novosibirsk State University, Novosibirsk 630090, Russia}
\author{Yu.V.~Usov}
\affiliation{Budker Institute of Nuclear Physics, SB RAS, Novosibirsk, 630090, Russia}
\affiliation{Novosibirsk State University, Novosibirsk 630090, Russia}
\author{A.V.~Vasiljev}
\affiliation{Budker Institute of Nuclear Physics, SB RAS, Novosibirsk, 630090, Russia}
\affiliation{Novosibirsk State University, Novosibirsk 630090, Russia}
\author{I.M.~Zemlyansky}
\affiliation{Budker Institute of Nuclear Physics, SB RAS, Novosibirsk, 630090, Russia}
\affiliation{Novosibirsk State University, Novosibirsk 630090, Russia}
\collaboration{The SND Collaboration}

\begin{abstract}
A search for the process $e^+e^- \to \eta^\prime$ has been performed
with the SND detector at the VEPP-2000 $e^+e^-$ collider.
The data were accumulated at the center-of-mass energy of $957.78\pm 0.06$
MeV with an integrated luminosity of about 2.9 pb$^{-1}$.
For reconstruction of the $\eta^\prime$ meson five decay chains
have been used:
$\eta^{\prime}\to\eta\pi^+\pi^-$ followed by the $\eta$ decays
to $\gamma\gamma$ and $3\pi^0 $,
and $\eta^{\prime} \to \eta\pi^0\pi^0$ followed by the $\eta$ decays
to $\pi^+\pi^-\pi^0$, $\gamma\gamma$, and $3\pi^0$.
As a result, the upper limit has been set on the $\eta^\prime$ electronic
width: $\Gamma_{\eta^{\prime}\to e^+e^-} < 0.0020$ eV at
the 90\% confidence level.
\end{abstract}


\maketitle

\section{Introduction}
\begin{figure}
\includegraphics[width=0.5\textwidth]{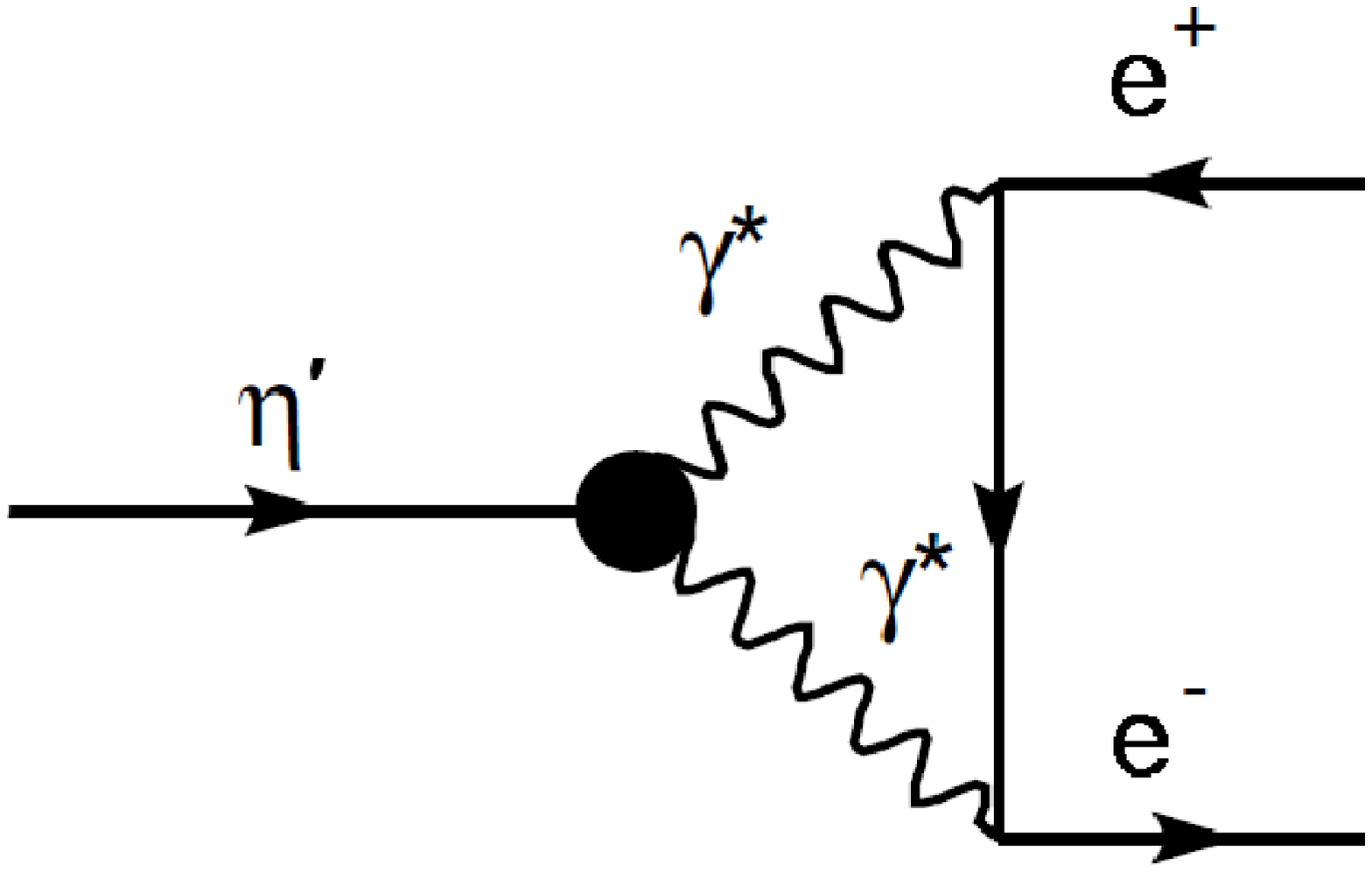}
\caption{The diagram for the $\eta^{\prime}\to e^+e^-$ decay.
\label{fig1}}
\end{figure}
This article is devoted to a search for the rare leptonic decay
$\eta^{\prime}\to e^+e^-$. In the Standard model this decay
proceeds through the two-photon intermediate state (Fig.~\ref{fig1})
and, therefore, is suppressed as $\alpha^2$ relative to the
$\eta^{\prime}$ two-photon decay.
An additional suppression of $(m_e/m_\eta^{\prime})^2$ arises from
helicity conservation. The imaginary part of the $ \eta^{\prime} \to 
e^+ e^-$ decay amplitude can be expressed in terms of the
known two-photon width $\Gamma(\eta^{\prime}\to \gamma\gamma)$.
Neglecting the real part of the amplitude  one can  obtain the
model-independent lower limit (unitary limit) on the decay probability
${\cal B}(\eta^{\prime}\to e^+e^-) > 3.8\times 10^{-11}$~\cite{unlim}.
Calculation of the real part requires knowledge of the transition form factor
$F(q_1^2,q_2^2)$ for the $\gamma^\ast\gamma^\ast\to\eta^{\prime}$ vertex,
where $q_1^2$ and $q_2^2$ are the photon virtualities in the loop.
The real part may increase the decay probability by a factor of 3--5
as compared with the unitary limit~\cite{th1,th2}.  Due to
the small probability, the $\eta^{\prime}\to e^+e^-$ decay may be
sensitive to contributions
not described by the Standard Model~\cite{newphys1,newphys2}.

The strictest limit on the decay branching fraction
${\cal B}(\eta^{\prime}\to e^+e^-) < 1.2\times 10^{-8}$~\cite{cmd} 
at the 90\% confidence level (CL) was
set recently in the experiments with the CMD-3 detector at the VEPP-2000
$e^+e^-$ collider~\cite{vepp2000}. In this experiment the technique of
using the inverse reaction $e^+e^-\to\eta^{\prime}$ for a 
measurement of ${\cal B}(\eta^{\prime}\to e^+e^-)$ proposed in Ref.~\cite{nd}
was applied. The cross section of the $e^+e^-\to\eta^{\prime}$  reaction
at the center-of-mass (c.m.) energy $E=m_{\eta^\prime}c^2$ is equal to
\begin{equation}
\sigma_0=\frac{4\pi}{m_{\eta^{\prime}}^2}B(\eta^{\prime}\to e^+e^-).
\label{xs0}
\end{equation}

In this paper we present the results of the search for the $\eta^{\prime}\to
e^+e^-$ decay in the experiments with the SND detector at the
VEPP-2000 collider. The SND data used in this analysis were collected
simultaneously with the CMD-3 data mentioned above.

\section{Detector and experiment}
The SND detector is described in detail elsewhere~\cite{SND}.
This is a nonmagnetic detector, the main part of which is a three-layer
spherical electromagnetic calorimeter based on NaI(Tl) crystals.
The solid angle covered by the calorimeter is 90\% of $4\pi$. Its energy
resolution for photons is $\sigma_E/E=4.2\%/\sqrt[4]{E({\rm GeV})}$, and
the angular resolution is about $1.5^\circ$. The directions of charged
particles are measured by a tracking system, which consists of the 9-layer
drift chamber and the proportional chamber with readout from cathode strips.
The tracking system covers a solid angle of 94\% of 4$\pi$. The
calorimeter is surrounded by a muon system, which is used, in particular,
for cosmic-background suppression.

Data used in this analysis with an integrated luminosity of
about 2.9 pb$^{-1}$ were accumulated in 2013 at the c.m. energy
close to $m_{\eta^{\prime}}c^2=957.78\pm0.06$ MeV~\cite{pdg}.
During the data taking period the beam energy was monitored with an
absolute accuracy of about 60 keV by
the Back-scattering-laser-light system~\cite{emes}.
The data taking conditions are described in detail in Ref.~\cite{cmd}.
The average value of the c.m. energy is $E_{\rm cm}=957.68\pm0.060$ MeV;
its spread is $\sigma_{E_{\rm cm}}=0.246\pm0.030$ MeV. To obtain
the cross section of $\eta^{\prime}$ production in the real experimental
conditions we have to take into account the radiative corrections to the
initial state, and the energy spread. This was done in Ref.~\cite{cmd}.
As the collider energy spread (FWHM = 0.590 MeV) is significantly larger
than the $\eta^{\prime}$ width $\Gamma_{\eta^{\prime}}=
(0.198\pm0.009)$ MeV~\cite{pdg}, the resulting cross section is proportional
to the electronic width
\begin{equation}
\sigma_{\rm vis}({\rm nb})=
(6.38\pm0.23)\Gamma_{\eta^{\prime}\to e^+e^-}({\rm eV}). 
\label{xsvis}
\end{equation}
It should be noted that the radiative corrections and the energy spread
lead to a reduction of the cross section compared to the Born one 
(Eq.~(\ref{xs0})) by a factor of four.

The search for the process $e^+e^-\to\eta^{\prime}$ is performed in 
five decay chains: $\eta^{\prime} \to \eta\pi^+\pi^-$ with the $\eta$ decays
to $\gamma\gamma$ and $3\pi^0$, and $\eta^{\prime} \to \eta\pi^0\pi^0$
with the $\eta$ decays to $\pi^+\pi^-\pi^0$, $\gamma\gamma$ and 
$3\pi^0$. For luminosity normalization of events with charged particles 
in the final state the large-angle Bhabha scattering
is used, while for events containing only photons the luminosity
is measured using the two-photon annihilation $e^+e^-\to \gamma\gamma$ . 
The corresponding integrated luminosities are measured  
to be $L_{ee}=2.91$ pb$^{-1}$ and $L_{\gamma\gamma}=2.82$ pb$^{-1}$. 
The difference between these values, about 3\%, gives us
a conservative estimate of the luminosity systematic uncertainty.

Using different normalizations allows to partly cancel  
systematic uncertainties associated with hardware event selection, 
charged track reconstruction, and beam-generated extra tracks.

\section{Event selection}
\subsection{\label{dc1}\bf\boldmath Decay chain $\eta^{\prime}\to\pi^+\pi^-\eta$, $\eta\to \gamma\gamma$ }
This $\eta^{\prime}$ decay channel has the largest probability,
about 17\%, and the lowest multiplicity among the channels studied
in this work. Because of the small multiplicity the background for this 
channel arises from almost all $e^+e^-$ annihilation processes.
In background processes with a small number of photons, as $e^+e^-\to e^+e^-(\gamma)$ or
$e^+e^-\to \pi^+\pi^-(\gamma)$, additional fake photons appear
as a result of splitting of electromagnetic showers, nuclear interaction
of pions in the calorimeter, or superimposing beam-generated background.
\begin{figure}
\includegraphics[width=0.4\textwidth]{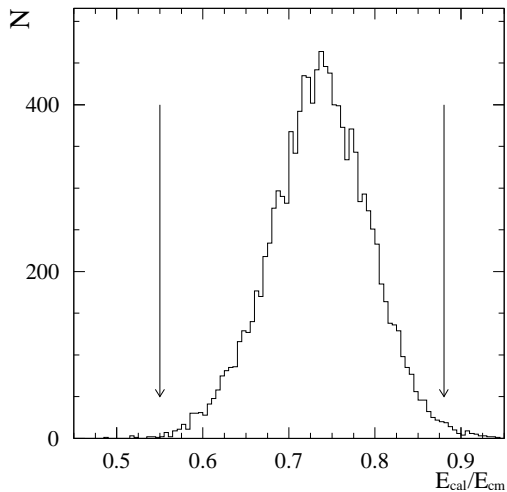}
\caption{The spectrum of the normalized total energy deposition in 
the calorimeter for simulated events of the process
$e^+e^-\to\eta^{\prime}\to\pi^+\pi^-\eta$, $\eta\to \gamma\gamma$.
The arrows indicate the boundaries of the selection cut
$0.55<E_{\rm cal}/E_{\rm cm}<0.9$.
\label{fig2}}
\end{figure}

At the first stage events with two charged particles originating
from the interaction region and two photons are selected. 
The muon system veto is applied to reject cosmic-ray background.
The charged particle tracks are fitted into a common vertex. 
Their polar angles must be in the range
$40^\circ<\theta<140^\circ$. To suppress background from collinear
two-body processes, mainly from $e^+e^-\to e^+e^-$, 
the azimuthal angles of the charged particles are required to
satisfy the condition $|180^\circ-|\phi_1-\phi_2||>10^\circ$. 
The background from $e^+e^-\to n\gamma$ events with photon conversion  
into an $e^+e^-$-pair is rejected by the condition $\psi_{cc}>20^\circ$, where
$\psi_{cc}$ is the open angle between the charged particles. To remove
events with fake photons from pion nuclear interactions in the calorimeter,
the condition on the minimal open angle between a charged particle and photon  
$\psi_{c\gamma}>20^\circ$ is applied.
The specific feature of this channel is a large energy deposition in 
the calorimeter $E_{\rm cal}$. The distribution of this parameter for 
simulated events of the process
$e^+e^-\to \eta^\prime \to \pi^+\pi^-\eta\to \pi^+\pi^-2\gamma$ is
shown in Fig.~\ref{fig2}. The arrows indicate the boundaries of the 
condition used $0.55<E_{\rm cal}/E_{\rm cm}<0.9$.
\begin{figure}
\includegraphics[width=0.6\textwidth]{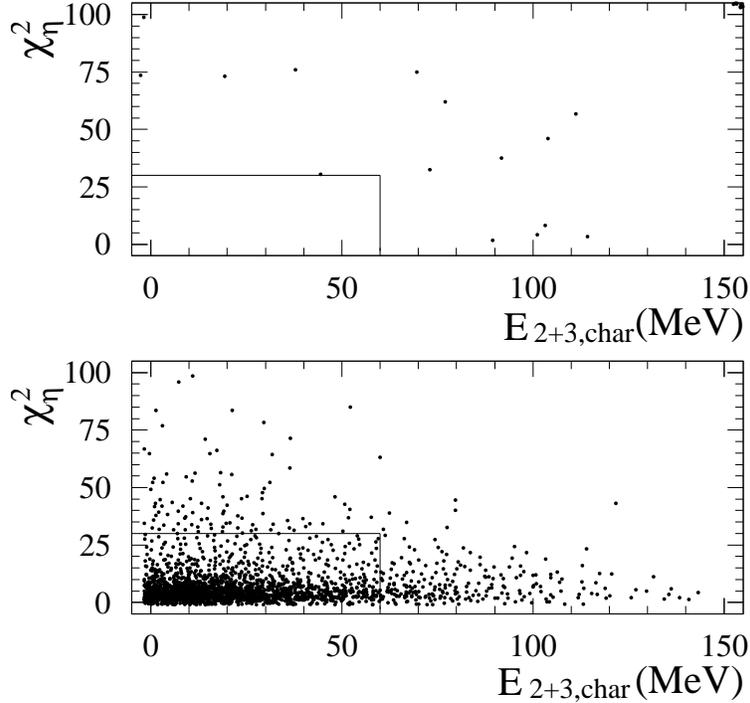}
\caption{ Two-dimensional  distribution of parameter $\chi^2_\eta$
over $E_{\rm char,2+3}$ for data events (top) and simulated events of the 
$e^+e^-\to\eta^{\prime}\to\pi^+\pi^-\eta$, $\eta\to \gamma\gamma$ process
(bottom). The rectangle in the bottom left corner of the plot corresponds to
the selection criteria $\chi^2_\eta<30$ and $E_{\rm char,2+3}<60$ MeV.
\label{fig3}}
\end{figure}

For events passing preliminary selection  
the kinematic fit to the $e^+e^-\to\pi^+\pi^-\eta$ hypothesis 
is performed. The quality of the fit is characterized by the 
parameter $\chi^2_\eta$.
Another important parameter used for the final selection is the
sum of energy depositions of charged particles in the second and third layers 
of the calorimeter $E_{\rm 2+3,char}$. Since pions in the process under study
are soft, they stop predominantly in the first layer of the calorimeter. 
The two-dimensional  distributions of the parameters $\chi^2_\eta$ and
$E_{\rm 2+3,char}$  for data events and simulated events of the
process under study are shown at Fig.~\ref{fig3}. The rectangle in the
bottom left corner corresponds to the selection criteria applied: 
$\chi^2_\eta<30$ and $E_{\rm char,2+3}<60$ MeV.

No data events are selected with the selection criteria described above. 
The detection efficiency for 
$e^+e^-\to\eta^{\prime}\to\pi^+\pi^-\eta, \eta\to 2\gamma$ events is
determined using Monte Carlo (MC) simulation to be $(12.2\pm 1.2)\%$. 
The quoted error is systematic. For its estimation we use the results of
the study of data-MC simulation difference in the measurement of 
the $e^+e^-\to \pi^+\pi^-\eta$ cross section~\cite{etapipi} at 
$E_{\rm cm}> 1.2$ GeV. 

The dominant sources of background for this decay mode after 
applying the selection criteria are the processes 
$e^+e^-\to \eta\gamma, \eta\to \pi^+\pi^-\pi^0$ and
$e^+e^-\to \pi^+\pi^-\pi^0\pi^0$. The number
of background events is estimated using MC simulation
to be $0.7\pm 0.1$ and
$0.10\pm 0.05$ for the first and second processes, respectively.
The values of the $e^+e^-\to \eta\gamma$ and 
$e^+e^-\to \pi^+\pi^-\pi^0\pi^0$ cross sections were taken from the
measurements~\cite{etagcmd,etagsnd,4pisnd}. 
The background can be also estimated from the two-dimensional distribution
shown in Fig.~\ref{fig3} using the assumption that the
$\chi^2_\eta$ and $E_{\rm 2+3,char}$ distribution are independent.
The number of background events in the signal rectangle 
($\chi^2_\eta<30, E_{\rm 2+3,char}<60$ MeV) is estimated as 
$n_2n_3/n_4\approx1\pm 1$, where $n_2$, $n_3$ and $n_4$ are the numbers of
data events in the regions ($30<\chi^2_\eta<60, E_{\rm 2+3,char}<60$ MeV),
($\chi^2_\eta<30, E_{\rm 2+3,char}>60$ MeV) and
($30<\chi^2_\eta<60, E_{\rm 2+3,char}>60$  MeV), respectively.

There is also the nonresonant reaction $e^+e^-\to \pi^+\pi^-\eta$ having
the same final state as the process under study. This reaction proceeds
through the $\rho\eta$ intermediate state and, therefore, is 
suppressed due to the small phase space of the final particles.
Interpolating the result of the fit to the 
$e^+e^-\to \pi^+\pi^-\eta$ cross section measured at higher energies
~\cite{etapipi}, we estimate that the nonresonant cross section
at $E_{\rm cm}=960$ MeV is about 1.7 pb. Assuming the same 
detection efficiencies for the resonant and nonresonant processes, the
nonresonant contribution is estimated to be 0.2 events.

\subsection{\bf\boldmath Decay chain $\eta^{\prime}\to\pi^+\pi^-\eta$, $\eta\to 3\pi^0$} 
In this decay mode the following selection criteria are used.
An event must contain two charged particles originating from the 
interaction region and six photons. We require the muon-system veto,
$|180-|\phi_1-\phi_2||>10^\circ$, $\psi_{c\gamma}>20^\circ$,
$0.5<E_{\rm cal}/E_{\rm cm}<0.9$, and $E_{\rm 2+3,char}<90$ MeV.

\begin{figure}
\includegraphics[width=0.65\textwidth]{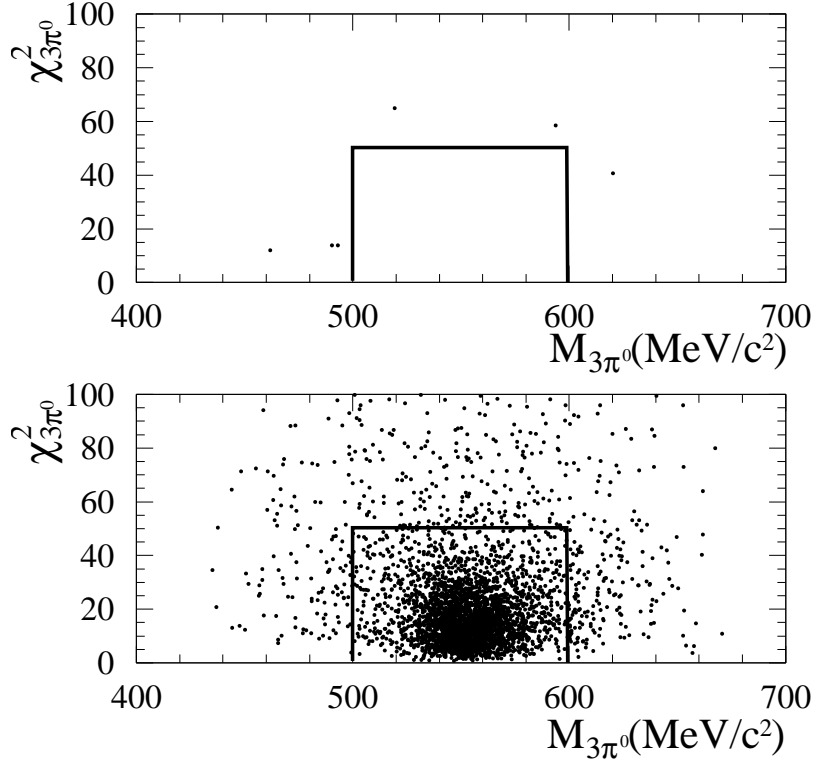}
\caption{The two-dimensional distribution of the parameters $\chi^2_{3\pi^0}$
and $M_{3\pi^0}$ for data events (top) and simulated  
$\eta^{\prime}\to\pi^+\pi^-\eta, \eta\to 3\pi^0$ events (bottom).
The rectangle corresponds to the selection criteria used:
$\chi^2_{3\pi^0}<50$ and $500<M_{3\pi^0}<600$ MeV/$c^2$.
\label{fig4}}
\end{figure}

For selected events the kinematic fit is performed to the
hypothesis $e^+e^-\to\pi^+\pi^-3\pi^0$.
The two-dimensional distributions of $\chi^2$ of the kinematic
fit ($\chi^2_{3\pi^0}$) versus the three $\pi^0$ invariant mass 
($M_{3\pi^0}$) for data events and simulated
events of $e^+e^-\to\eta^{\prime}\to\pi^+\pi^-\eta$, $\eta\to 3\pi^0$
process are shown in Fig.~\ref{fig4}.
The following cuts on these parameters are used:
$\chi^2_{3\pi^0}<50$ and $500<M_{3\pi^0}<600$ MeV/$c^2$.
No data events satisfying the selection criteria applied are found.
The detection efficiency for  
$e^+e^-\to\eta^{\prime}\to\pi^+\pi^-\eta, \eta\to 3\pi^0$ events
determined using MC simulation is $(7.5\pm 0.8)\%$. The quoted error
is estimated according to Ref.~\cite{etapipi}. 

\begin{figure}
\includegraphics[width=0.4\textwidth]{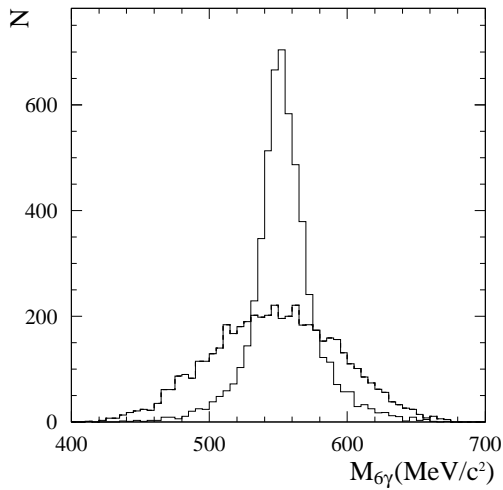}
\caption{The distribution of the three-$\pi^0$ invariant mass
for the decay channels:
$\eta^{\prime}\to\pi^+\pi^-\eta, \eta\to 3\pi^0$ (narrow distribution) and
$\eta^{\prime}\to\pi^0\pi^0\eta, \eta\to \pi^+\pi^-\pi^0$
(wide distribution), obtained using MC simulation.
\label{fig5}}
\end{figure}

It is necessary to note that the same final state $\pi^+\pi^-3\pi^0$ can be
obtained in the other decay chain 
$\eta^{\prime}\to\pi^0\pi^0\eta, \eta\to \pi^+\pi^-\pi^0$.
The distribution of the three-$\pi^0$ invariant mass for this decay
channel is shown in Fig.~\ref{fig5}.

It is seen that a significant part of 
$\eta^{\prime}\to\pi^0\pi^0\eta, \eta\to \pi^+\pi^-\pi^0$
events satisfies the condition  $500<M_{3\pi^0}<600$ MeV/$c^2$. 
The detection efficiency for this channel calculated 
using MC simulation is $(4.9\pm0.5)\%$.
We could not increase the detection efficiency for  
$\eta^{\prime}\to\pi^0\pi^0\eta, \eta\to \pi^+\pi^-\pi^0$ events
by using conditions on parameters specific for this decay mode,
for example, the $\pi^+\pi^-\pi^0$ invariant mass
instead of $M_{3\pi^0}$.

The dominant background source for the $\pi^+\pi^-\pi^0\pi^0\pi^0$
final state is the process $e^+e^-\to \pi^+\pi^-\pi^0\pi^0$.
Additional fake photons can appear as a result of nuclear interaction 
of charged pions or beam background. The number of background events 
obtained using MC simulation is $2.7\pm 0.5$. Since appearance of two 
fake photons is needed, the simulation can be used only for rough estimation 
of the background level. The background can be also estimated using
the $M_{3\pi^0}$ distribution for data. Based on four observed data events with  
$\chi^2_{3\pi^0}<50$ in Fig.~\ref{fig4}  and assuming a linear
background $M_{3\pi^0}$ distribution,
we estimate the background in the interval $500<M_{3\pi^0}<600$
MeV/$c^2$ to be  $2\pm 1$ events.
The nonresonant background from the $e^+e^-\to \pi^+\pi^-\eta$ process
discussed above in Sec.~\ref{dc1} is about 0.1 events in this decay mode.

\subsection{\bf\boldmath Decay chain $\eta^{\prime}\to\pi^0\pi^0\eta$, $\eta\to \gamma\gamma$ }
The event selection in this decay mode is performed in two steps.
At the first stage six-photon events containing no tracks in 
the drift chamber are selected. Each photon is required 
to have the transverse energy distribution in the calorimeter consistent with
the distribution for an electromagnetic shower~\cite{xinm}. The total energy 
deposition $E_{\rm cal}$ and the event momentum $P_{\rm cal}$ calculated 
using energy depositions in the calorimeter crystals must satisfy the 
following conditions:
\begin{equation}
0.7 < E_{\rm cal}/E_{\rm cm} < 1.2,~cP_{\rm cal}/E_{\rm cm} < 0.3,~
E_{\rm cal}/E_{\rm cm} - cP_{\rm cal}/E_{\rm cm} > 0.7.
\label{eton_vs_ptrt}
\end{equation}
To reject cosmic-ray background the muon-system veto is required.

\begin{figure}
\includegraphics[width=0.4\textwidth]{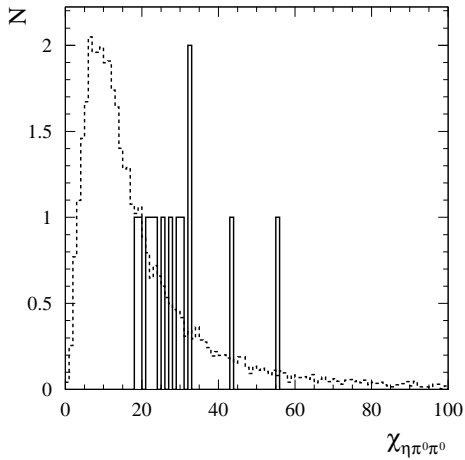}
\caption{The $\chi^2_{\eta\pi^0\pi^0}$ distribution
for data events (solid histogram) and simulated 
$e^+e^-\to\eta^\prime \to 2\pi^0\eta \to 6\gamma$ events
(dashed histogram).
\label{fig6}}
\end{figure}

For events passing initial selection the kinematic fit to the 
$e^+e^-\to\eta^{\prime}\to\eta\pi^0\pi^0\to 6\gamma$ hypothesis is
performed. The quality of the fit is characterized by 
the parameter $\chi^2_{\eta\pi^0\pi^0}$. The distributions of this parameter
for data events and simulated events of the 
$e^+e^-\to\eta^\prime \to \pi^0\pi^0\eta \to 6\gamma$ process are shown
in Fig.~\ref{fig6}.
The condition $\chi^2_{\eta\pi^0\pi^0} < 15$ is applied.

No data events satisfying the criteria described above 
have been found. The detection efficiency for the 
$e^+e^-\to\eta^\prime \to \pi^0\pi^0\eta$, $\eta\to \gamma\gamma$ process
obtained using MC simulation is $(14.6\pm 0.7)\%$. The quoted 
error is systematic, estimated using our work~\cite{SNDomegapi} on
the measurement of the $e^+e^-\to \pi^0\pi^0\gamma$ cross section.

The main background sources for this decay mode 
are the processes $e^+e^-\to\eta\gamma\to 3\pi^0\gamma$ and
$e^+e^-\to \pi^0\pi^0\gamma$.  Their cross sections were measured
in Refs.~\cite{etagcmd,etagsnd,pipigsnd,pipigcmd} and in this paper
(see Sec.~\ref{ul}). The number of background events from these sources is 
calculated to be $1.3\pm 0.3$ and $0.4\pm 0.1$, respectively. 
It should be noted that the number
of data events with $\chi^2_{\eta\pi^0\pi^0} < 100$ equal to 13
is in good agreement with the background prediction based on MC simulation:
$12\pm 2$ for $e^+e^-\to\eta\gamma$ and $3\pm 1$ for
$\pi^0\pi^0\gamma$.

\subsection{\bf\boldmath Decay chain $\eta^{\prime}\to\pi^0\pi^0\eta$, $\eta\to 3\pi^0$ }
For this decay mode with ten photons in the final state there is no
background from $e^+e^-$ annihilation. The main source of background is
cosmic-ray showers. We select events containing nine or more photons
and no tracks in the drift chamber. 
The photons must have the transverse energy distribution in the calorimeter 
consistent with  the distribution for an electromagnetic shower. 
The parameters $E_{\rm cal}$ and $P_{\rm cal}$ must satisfy the 
conditions (\ref{eton_vs_ptrt}). The muon system signal is required
to suppress cosmic-ray background. No data events are selected
after applying  these criteria.
The detection efficiency  for this decay mode is $(22.6\pm 1.1)\%$.

\section{Upper limit for $\eta^{\prime}\to e^+e^-$ decay \label{ul}}
The visible cross section for the process  $e^+e^- \to \eta^{\prime}$
is calculated as follows 
\begin{equation}
\sigma_{\rm vis}^{\rm exp}=\frac{N_s}{\sum{L_i\varepsilon_i}},
\label{sig1}
\end{equation}
where  $N_s$ is the sum of experimental events selected in the five 
decay modes, $\varepsilon_i$  is the detection efficiency in the mode $i$,
which includes the branching fractions for the corresponding 
$\eta^{\prime}$ and $\eta$ decays. The integrated luminosity $L_i$ 
is equal to $L_{ee}=2.91$ pb$^{-1}$ for the decay modes with charged particles
and  $L_{\gamma\gamma}=2.82$ pb$^{-1}$ for the multiphoton modes.
The denominator in the formula (\ref{sig1}) can be represented as 
$L_{ee}\varepsilon_s$. For the selection criteria described in the previous
section $\varepsilon_s=(6.2\pm 0.4)\%$.
Since the number of selected data events is equal to zero, 
we set the upper limit on the cross section. 
The technique of Cousins and Highland~\cite{Cousins} following 
the implementation of Barlow~\cite{Barlow} is used to calculate the  
limit with all uncertainties included ($N_s<2.32$ for 90\% CL): 
\begin{equation}
\sigma_{\rm vis}^{\rm exp}<12.7\mbox{ pb at 90\% CL.}
\end{equation}
The limit on the cross section is translated using Eq.(\ref{xsvis}) to
the upper limit on the $\eta^{\prime}$ electronic width  
\begin{equation}
\Gamma_{\eta^{\prime}\to e^+e^-}< 0.0020\mbox{eV at 90\% CL.} 
\end{equation}

As a test, we perform measurements of
the cross sections for the processes
$e^+e^-\to \pi^+\pi^-\pi^0$, $e^+e^- \to \pi^0\pi^0\gamma$ and
$e^+e^-\to \eta\gamma \to 3\pi^0\gamma$.
The process $e^+e^-\to \eta\gamma$ was studied in the seven-photon
final state. The events of these processes are selected with 
criteria similar to those described in the previous section. 
The obtained Born cross sections
$\sigma(e^+e^-\to \pi^+\pi^-\pi^0)=11.7\pm 0.2$ nb,
$\sigma(e^+e^-\to \pi^0\pi^0\gamma)=285\pm 21$ pb,
$\sigma(e^+e^-\to \eta\gamma)=244\pm 30$ pb
are in good agreement with the results of the previous measurements
$11.33\pm0.64$ nb~\cite{3pisnd} for $e^+e^-\to \pi^+\pi^-\pi^0$,
$242^{+89}_{-67}$ pb~\cite{pipigsnd} and
$390^{+112}_{-98}$ pb~\cite{pipigcmd} for $e^+e^-\to \pi^0\pi^0\gamma$,
$300\pm110$ pb~\cite{etagcmd} and
$390^{+140}_{-110}$ pb~\cite{etagsnd} for $e^+e^-\to \eta\gamma$.

\section{Conclusion}
The search for the process $e^+e^-\to \eta^\prime$ has been
performed in the experiment with the SND detector 
at the VEPP-2000 $e^+e^-$-collider.
To reconstruct the $\eta^\prime$-meson the five decay chains have
been used:
$\eta^{\prime} \to \eta\pi^+\pi^-$ followed by the $\eta$ decays to
$\gamma\gamma$ and $3\pi^0$, and 
$\eta^{\prime} \to \eta\pi^0\pi^0$ with $\eta$ decays into 
$\pi^+\pi^-\pi^0$, $\gamma\gamma$, and $3\pi^0$. 
No data events of  the $e^+e^-\to \eta^\prime$ process have been found.
Since the visible cross section for the process under study is 
proportional to the $\eta^\prime$ electronic width, 
we set the upper limit 
\begin{equation}
\Gamma_{\eta^{\prime}\to e^+e^-}< 0.0020\mbox{eV at 90\% CL}. 
\end{equation}
The obtained limit is slightly better than the limit set recently 
in the CMD-3 experiment $\Gamma_{\eta^{\prime}\to e^+e^-}< 0.0024$
eV~\cite{cmd}. 

Using the formula (\ref{sig1}) we
combine the SND (0 events, $L_i=2.91$ pb$^{-1}$,
$\varepsilon_i=(6.2\pm0.4)\%$) and CMD-3 (0 events, $L_i=2.69$
pb$^{-1}$, $\varepsilon_i=(5.3\pm0.3)\%$) data and obtain the  
combined upper limits on the electronic width
\begin{equation}
\Gamma_{\eta^{\prime}\to e^+e^-}< 0.0011\mbox{ eV at 90\% CL}. 
\end{equation}
and the branching fraction
[$\Gamma_{\eta^{\prime}}=(0.198\pm0.009)$ MeV~\cite{pdg}]
\begin{equation}
{\cal B}(\eta^{\prime}\to e^+e^-) < 5.6\times10^{-9}
\mbox{ at 90\% CL}.
\end{equation}
The obtained upper limit is most stringent but still
30-50 times larger than theoretical
predictions~\cite{th1,th2} made in the framework of the Standard Model.

\section{ACKNOWLEDGMENTS}
We thank S.I. Eidelman for fruitful discussions. Part of this
work related to the photon reconstruction algorithm in the
electromagnetic calorimeter and analysis of multiphoton events is supported 
by Russian Science Foundation (project N 14-50-00080).
This work is partly supported by the RFBR grant No. 15-02-03391.

\end{document}